\documentclass[aps,prl,twocolumn,showpacs,floatfix,nofootinbib,preprintnumbers,superscriptaddress,amsmath,amssymb]{revtex4-1}

\usepackage{dcolumn}
\usepackage{bm}
\usepackage{graphicx}
\usepackage{epstopdf}
\usepackage{wrapfig}
\usepackage{array} 
\usepackage{listings}
\usepackage[para,online,flushleft]{threeparttablex}
\usepackage{booktabs,dcolumn}
\usepackage{hyperref}

\usepackage{CJKutf8}
\usepackage{textpos}
\usepackage{booktabs}
\usepackage{multirow,bigdelim}
\usepackage{float}
\usepackage{xcolor}

\begin{document}

\begin{CJK*}{UTF8}{gbsn}

\title{Neutron drip line in the Ca region from Bayesian model averaging}

\author{L{\'e}o Neufcourt}
\affiliation{Department of Statistics and Probability, Michigan State University, East Lansing, Michigan 48824, USA}
\affiliation{FRIB Laboratory,
Michigan State University, East Lansing, Michigan 48824, USA}

\author{Yuchen Cao (曹宇晨)}
\affiliation{Department of Physics and Astronomy and NSCL Laboratory,
Michigan State University, East Lansing, Michigan 48824, USA}

\author{Witold Nazarewicz}
\affiliation{Department of Physics and Astronomy and FRIB Laboratory,
Michigan State University, East Lansing, Michigan 48824, USA}

\author{Erik Olsen}
\affiliation{FRIB Laboratory,
Michigan State University, East Lansing, Michigan 48824, USA}

\author{Frederi Viens}
\affiliation{Department of Statistics and Probability, Michigan State University, East Lansing, Michigan 48824, USA}

\begin{abstract}
The region of  heavy calcium isotopes forms the frontier of experimental and theoretical nuclear structure research where the basic concepts of nuclear physics are put to stringent test. The recent discovery of the  extremely neutron-rich nuclei around $^{60}$Ca \cite{Tarasov2018}  and the experimental determination of masses for $^{55-57}$Ca \cite{Michimasa2018} provide unique information about the binding energy  surface in this  region. To assess the impact of these  experimental discoveries on the nuclear landscape's extent, we use global mass models and statistical machine learning to make predictions, with quantified levels of certainty, for bound nuclides between Si and Ti. Using a Bayesian model averaging analysis based on   Gaussian-process-based extrapolations  we introduce the posterior probability $p_{ex}$ for each nucleus to be bound to neutron emission. We find that extrapolations for drip-line locations, at which the nuclear binding ends,
are consistent across the global mass models used, in spite of significant variations between their raw predictions. In particular, considering the current experimental information and current global mass models, we predict that $^{68}$Ca has an average posterior probability ${p_{ex}\approx76}$\%  to be bound to two-neutron emission while the nucleus $^{61}$Ca is  likely to decay by emitting a neutron (${p_{ex}\approx 46}$\%).
\end{abstract}

\maketitle
\end{CJK*}

\textit{Introduction --} 
How many protons and neutrons can form a bound atomic nucleus? 
Out of about 3,200 isotopes  known \cite{Thoennessen2016}
only 286 primordial nuclides have existed in their current form since before Earth was formed. They form the valley of  stability on the nuclear landscape. 
Moving away from the region of stable isotopes by adding neutrons or protons, one enters the regime of short-lived radioactive  nuclei,  which are beta unstable.
Nuclear existence ends at the   ``drip lines", where the last nucleons  are no longer attached to the nucleus  by the strong interaction and  drip off. According to current theoretical estimates \cite{Erl12a,Agbemava2014} the number of
bound nuclides with atomic number $Z$ between 2 and 120  is around 7,000.

The particle stability of a nuclide is determined by its separation energy, i.e., 
the   energy required to remove from it a single nucleon or a pair of like nucleons.
If the separation energy is positive, the 
nucleus is bound to nucleon decay; if the separation
energy is negative, the nucleus is particle-unstable. In this Letter, we study the one-neutron ($S_{1n}$) and two-neutron ($S_{2n}$) separation energies of neutron-rich nuclei.
The drip line is reached when the  separation energy reaches zero; hence, one can talk about the one-neutron drip line when $S_{1n}=0$ and the two-neutron drip line when $S_{2n}=0$. 
Very weakly bound, or unbound,  nuclei that lie in the immediate vicinity of drip lines are referred to as threshold systems.
The separation energies and
 drip-line positions are strongly
 affected by nucleonic pairing, or nuclear superfluidity \cite{Brink2005}. Since it costs energy to break a nucleonic pair,  nuclei with even numbers of nucleons are more bound than their
 odd-nucleon-number neighbors. As a result, the one-nucleon drip line is reached earlier than the
two-nucleon drip line, which results in  a highly irregular
pattern of nuclear existence that meanders between odd- and even-particle systems.

The territory of neutron-rich nuclei is arguably the most fertile ground for breakthroughs  in  nuclear structure research and the Ca region is of particular interest. The heaviest Ca isotope discovered to-date is $^{60}$Ca \cite{Tarasov2018}. This nucleus, having $Z=20$ protons and $N=40$ neutrons, i.e., 
containing 12 more neutrons than the heaviest stable calcium isotope,
 was found recently together with seven other neutron-rich nuclei: $^{47}$P, $^{49}$S, $^{52}$Cl, $^{54}$Ar, $^{57}$K, $^{59}$Ca, and $^{62}$Sc.  In addition, one event consistent with 
$^{59}$K was registered \cite{Tarasov2018}. This discovery extends the range of known nuclei in this region, previously established in Refs.~\cite{Tarasov2009,Tarasov2013}. In  separate experimental studies, the atomic masses of 
$^{55-57}$Ca  were determined \cite{Michimasa2018} and the uncertainties of the $^{52-55}$Ti mass values were significantly  reduced \cite{Leistenschneider2018}.

The Ca region is arguably the most critical one to look at from a theory perspective, because it provides an exciting opportunity to bridge  the refined methods based on realistic interactions, in which all $A$ nucleons are considered as elementary degrees of freedom, with nuclear density functional theory (DFT) employing  energy density functionals (EDFs) expressed in terms of proton and neutron local densities and currents.
During recent years, the  $A$-body approaches reached {\it selected} medium-mass nuclei and provided  predictions for their global properties and spectroscopy \cite{Hagen2014,Hergert2016}. 
Nuclear DFT offers a more coarse-grained picture of nuclei than $A$-body approaches, but
can be applied globally  across the nuclear chart from light to superheavy nuclides \cite{Erl12a,Agbemava2014}.
The associated  EDFs  are primarily constrained by global nuclear
properties such as binding energies and radii \cite{Ben03}. By considering symmetry-breaking effects, nuclear DFT can describe on the same footing spherical nuclei close to magic shells and deformed open-shell systems.
A well-controlled link between $A$-body methods  and DFT is essential if one aims to understand nuclei and nucleonic matter from a bottom-up perspective \cite{Forssen2013,Nazarewicz2016}. 

When it comes to the Ca chain itself, $A$-body methods  provide an excellent description of binding energies, charge radii, and spectroscopy up to $^{54}$Ca, depending on the interaction used 
\cite{Hagen2012,Holt2014,Soma2014,Hergert2014,Hagen2015,GarciaRuiz16,Simonis2017,Stroberg2017}.
Likewise, DFT approaches with globally-optimized EDFs reproduce measured global properties. However, there is no consensus when it comes to extrapolations towards the neutron drip line. Namely, $A$-body methods with two- and three-body interactions predict the two-neutron drip-line around $^{60}$Ca\cite{Hagen2013,Stroberg2017} while the DFT approaches locate it around $^{70}$Ca\cite{Erl12a,Forssen2013}. 

In this Letter, we investigate what global nuclear mass models, aided by Bayesian machine learning, can tell us about the topography of the mass surface and neutron drip-lines in the Ca region. Our methodology roughly follows the recent paper \cite{Neufcourt2018}. 
Since
Bayesian machine learning  requires a sufficient number of data points in order to make extrapolations with reasonable certainty, one must work with models which are mostly global. To this end,
we consider global models based on nuclear DFT with realistic Skyrme EDFs as well as the more phenomenological mass models FRDM-2012 and HFB-24 rooted in the mean-field theory. 

\textit{Density functional theory calculations --} 
We used the  DFT mass predictions
based on SkM$^*$ \cite{Bartel1982}, SkP \cite{Dob84}, SLy4 \cite{Chabanat1995}, SV-min \cite{Kluepfel2009}, UNEDF0 \cite{UNEDF0},  and UNEDF1 \cite{UNEDF1}
EDFs stored in the theoretical database MassExplorer \cite{massexplorer}. 
The UNEDF2 \cite{UNEDF2} mass table has been computed exactly in the same way as in Ref.~\cite{Erl12a}.
The DFT predictions are compared to the results of the global mass models FRDM-2012 \cite{Moller2012} and HFB-24 \cite{Goriely2013}.

DFT calculations were carried out for even-even nuclei as
we want to avoid additional complications and uncertainties related to the
choice and treatment of quasi-particle configurations in odd-$A$ and odd-odd
systems \cite{Bonneau2007,Schunck2010,Afanasjev2015a}.
Binding energies of odd-$A$ and odd-odd nuclei were obtained from the binding energy values and average pairing gaps computed for even-even neighbors. The associated error on $S_{1n}$ is expected to be 200-300\,keV \cite{Schunck2010}. For completeness, in our analysis we also considered
even-even nuclei predicted to lie just beyond the two-neutron drip line, i.e., those with a slightly positive neutron chemical potential. 
Those results should be considered as rough estimates as the HFB theory does not guarantee that the nucleonic densities and fields are {localized} in this regime \cite{Dob84,Dob2013}.

\textit{Statistical analysis --}
We first compute the so-called separation energy residuals, i.e., the differences
$\delta_{1n/2n}(Z,N):=S_{1n/2n}^{\mathrm{exp}}(Z,N)-S_{1n/2n}^{\mathrm{th}}(Z,N)$,
between experimental values and model predictions of $S_{1n/2n}$, based on  the training  datasets AME2003 \cite{AME03b} and AME2016* consisting of  AME2016 masses \cite{AME16b} supplemented by  the recently updated 
$^{52-55}$Ti masses \cite{Leistenschneider2018}
(the subscripts $1n/2n$ are used to indicate either one-neutron or two-neutron separation energies).
Using the values of $\delta_{1n/2n}(Z,N)$ for those training nuclei $(Z,N)$, we construct emulators 
$\delta_{1n/2n}^{\mathrm{stat}}(Z,N)$
using a Bayesian machine learning analysis of extrapolations via Gaussian processes (GP)
following the methodology  previously developed in Ref.~\cite{Neufcourt2018}.
Our likelihood, the GP model, is a popular way \cite{RasmussenWilliams} of interpolating or extrapolating quantities from {neighboring} ones. 
 It is strongly based on the assumption of a local spatial structure in the data,  and contains required uncertainty {modeling}.
We took the GP model in a form of  a mean-zero Gaussian random field with a {quadratic} exponential spatial covariance kernel \cite{MacKay} featuring three parameters: its scale $\eta$, which represents a noise intensity, 
and two characteristic spatial lengths $\rho$: one in the proton direction and  one in the neutron direction; see Supplemental Material (SM) \cite{SM} for details. 
We performed the statistical analysis independently on the sets of $S_{1n}$ and $S_{2n}$, respectively for odd-$N$ and even-$N$ nuclei, and independently for odd-$Z$ and even-$Z$ nuclei.

Posterior samples are obtained via 100,000 iterations of the Metropolis \cite{Hastings1970} algorithm, from 
which the posterior mean value provides our predictions while
Bayesian credibility intervals (CIs) are built using the corresponding posterior quantiles, symmetric around the mean, 
at all uncertainty levels.
We also evaluate the performance of the prediction via a comparison of the rms
deviation before and after statistical refinement. 
It is worth noting that our statistical CI estimates take into account all possible sources of uncertainty including statistical, numerical, and systematic uncertainty, including model approximations and modeling uncertainty within the DFT framework.

The unknown separation energies are  predicted statistically by combining the theoretical predictions and the credibility intervals for the estimated residuals. For instance, the estimated prediction at $(Z,N)$ for the one-neutron (resp. two-neutron) separation energy is 
$S_{1n/2n}^{\mathrm{est}}(Z,N)=S_{1n/2n}^{\mathrm{th}}(Z,N,)+\delta_{1n/2n}^{\mathrm{stat}}(Z,N)$ where the last term is the Bayesian posterior mean prediction for the residual.
A similar strategy of correcting model predictions outside the training domain by estimated residuals  has recently  been applied in Ref.~\cite{Utama18}. 
(Applications of statistical methods to mass predictions have been described in several papers \cite{Athanassopoulos2004,Bayram2017,Yuan2016,Utama16,Utama17,Bertsch2017,Zhang2017,Niu2018},  primarily in the context of interpolations.)
The new aspect of our work lies in that we apply
the Bayesian method to provide a full quantification of the uncertainty surrounding the point estimate. For more details we refer the reader to SM.

\begin{figure}[htb]
\includegraphics[width=0.9\linewidth]{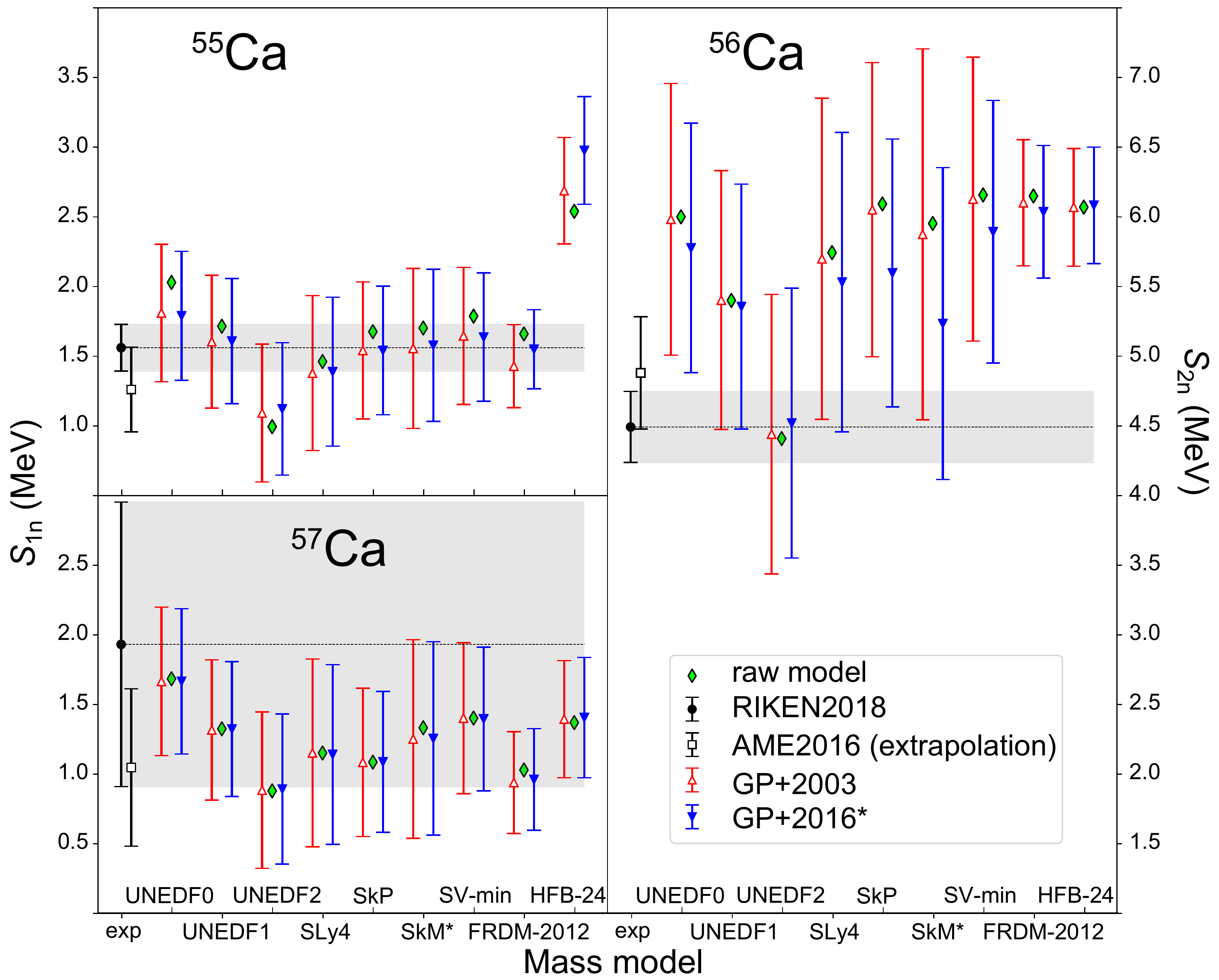}
\caption{\label{fig-Ca57} 
One-neutron separation energy for $^{55,57}$Ca (left) and 
two-neutron separation energy for $^{56}$Ca (right) calculated with the nine global mass models with statistical correction obtained with GP trained on the AME2003 (GP+2003) and  AME2016* datasets. The recent data from Ref.~\cite{Michimasa2018} (RIKEN2018) and the extrapolated AME2016 values \cite{AME16b} are marked.
The shaded regions are one-sigma error bars from Ref.\cite{Michimasa2018}; error bars on theoretical results are one-sigma credible intervals computed with GP extrapolation.}
\end{figure}
\textit{Results --} GP's superior predictive power was assessed in Ref.~\cite{Neufcourt2018} for the $S_{2n}$  of even-even nuclei.  The present work achieves comparable performances for odd-$Z$ nuclei and for $S_{1n}$ values, with prediction improvements ranging from 20 \% to 40 \% for most models (see SM).
To further assess the performance of our approach, we apply it to the recently measured masses of  $^{55-57}$Ca \cite{Michimasa2018}.  As seen in Fig.~\ref{fig-Ca57},
the predicted $S_{1n}$ values for $^{55,57}$Ca are consistent with experiment for most models while the $S_{2n}$ of $^{56}$Ca is  slightly overestimated. 
The impact of newer mass measurements beyond AME2003 on our predictions is minor; this is because very few datapoints that can impact our local GP model were added in the Ca region.  The large deviation in the $S_{1n}$ of $^{55}$Ca in HFB-24 is noteworthy. As illustrated in SM and  Ref.~\cite{Neufcourt2018}, neutron separation energies predicted by this model often exhibit irregular behavior.

\begin{figure}[htb]
\includegraphics[width=0.9\linewidth]{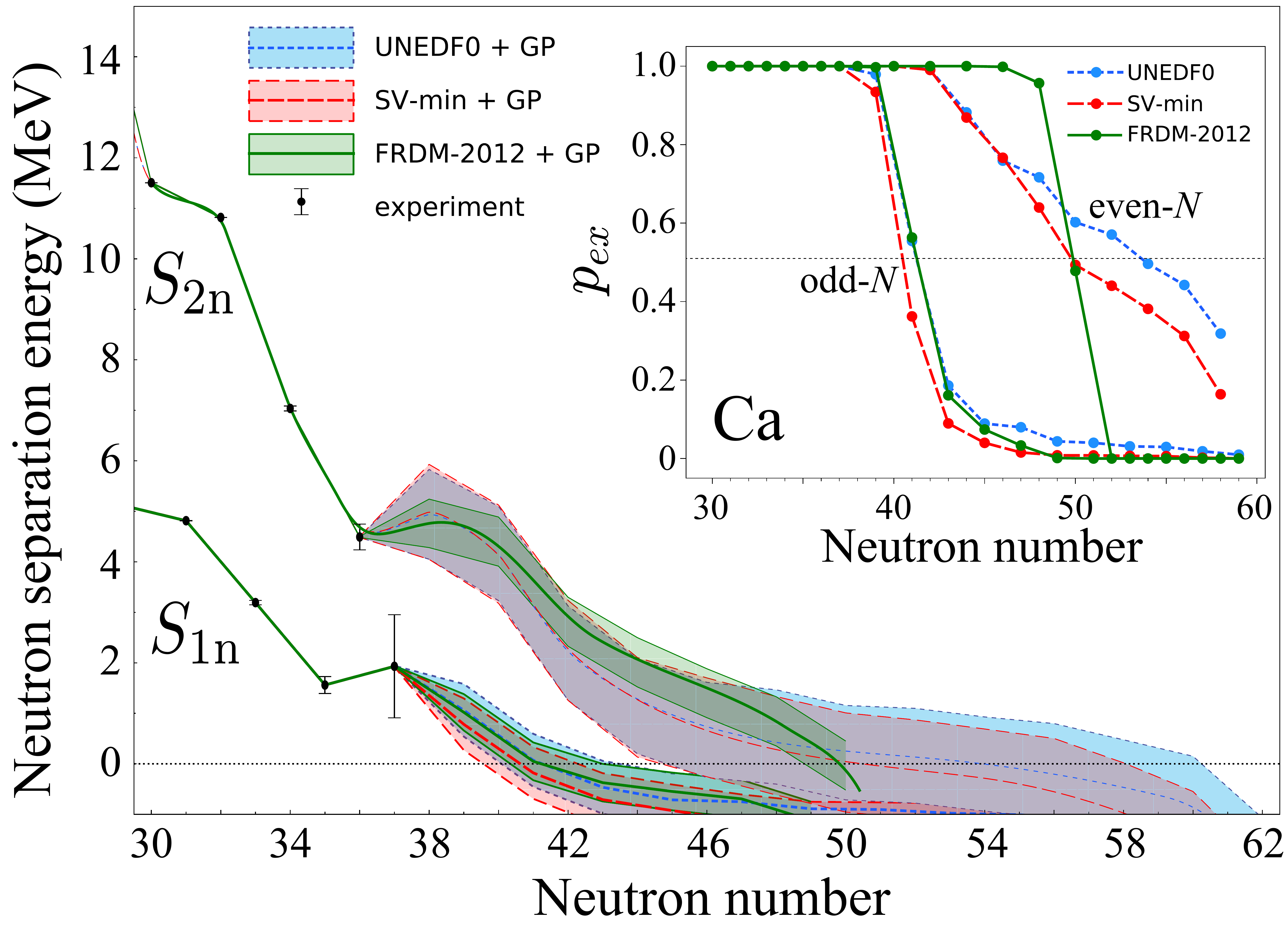}
\caption{\label{fig-Cachain}
Extrapolations of $S_{1n}$ and  $S_{2n}$ for the Ca chain corrected with GP and one-sigma CIs, combined for three
representative models. 
The solid lines show the average prediction while the shaded bands give one-sigma CIs.
The insert  shows the posterior probability of existence for the Ca chain. The $p_{ex}=0.5$ limit is marked by a dotted line. For the Ti-chain plot, see
SM \cite{SM}. 
}
\end{figure}

\begin{figure*}[htb]
\includegraphics[width=0.9\linewidth]{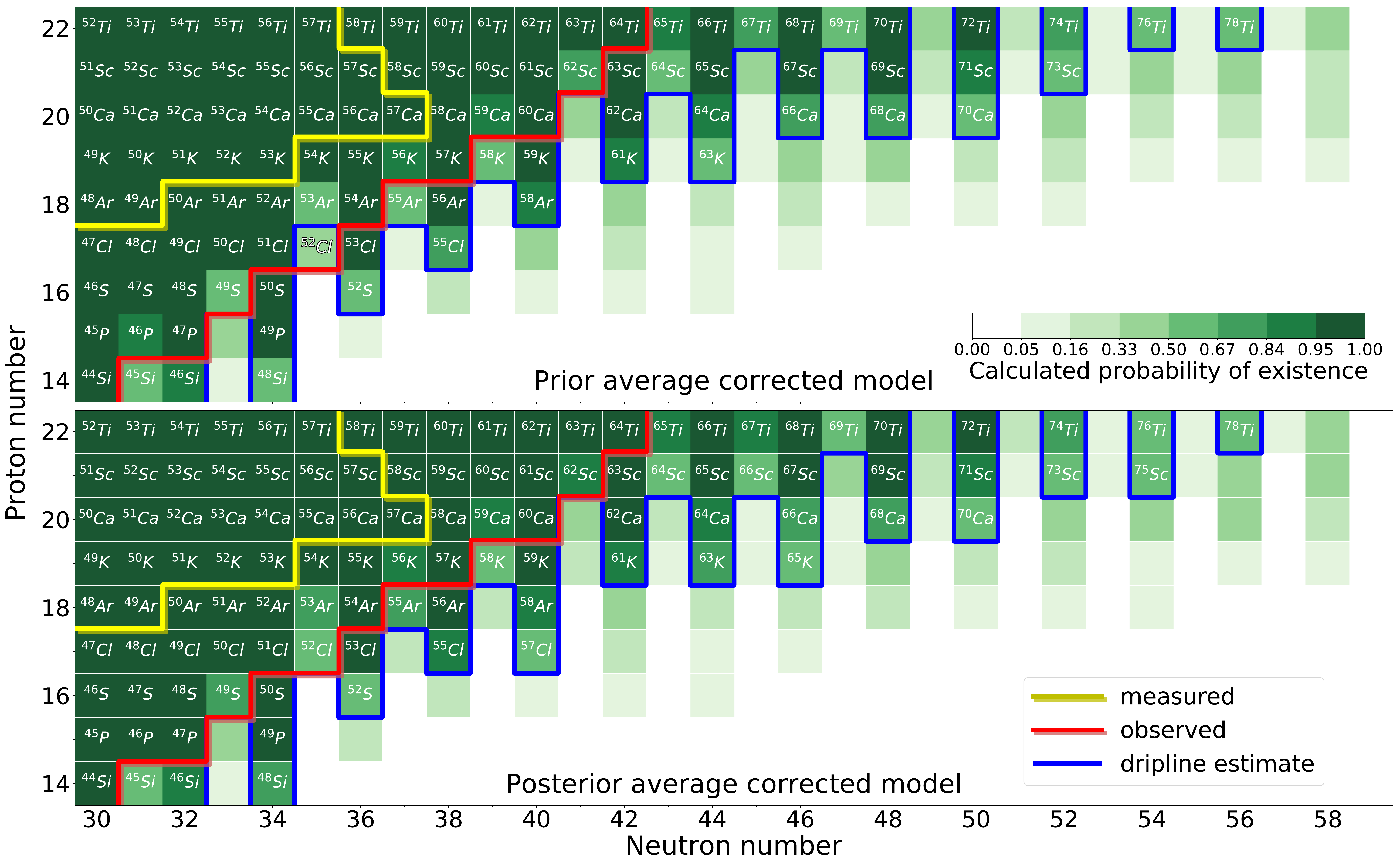}%
\caption{
\label{fig-dripline} Posterior probability of existence of neutron-rich nuclei in the Ca region averaged over all  models. Top:  uniform model averaging. Bottom:
averaging using 
posterior weights \eqref{eq-post-weights} constrained by the existence of  $^{52}$Cl,  $^{53}$Ar, and $^{49}$S.
The range of nuclei with experimentally-known masses is marked by a yellow line. The red line marks the limit of nuclei 
that have been  experimentally observed. The estimated drip line that separates  the $p_{ex} > 0.5$ and $p_{ex}<0.5$
regions is indicated by a blue line. 
}
\end{figure*}

Figure~\ref{fig-Cachain}  shows  extrapolated separation energies  for the Ca  isotopic chain  for three global mass models corrected with the GP emulator.
(Here and in the following we  shall use the notation ``model+GP" (e.g., UNEDF0+GP) to emphasize that the statistical corrections are done with the GP emulator.) 
The models are consistent overall once the statistical correction and uncertainty are taken into  account.  According to the computed empirical coverage probabilities \cite{Gneiting2007,Gneiting2007a}, our credibility intervals are slightly conservative for large credibility levels (see Sec. I.C of SM for more discussion).

For a given isotopic chain and nuclear model,
one obtains an upper bound on the location of the first isotope at which the binding energy becomes negative, depending on the choice of credibility level. For instance, the posterior mean values (full lines) of the
UNEDF0+GP model place the $2n$ drip line for Ca around $N=54$, 
while considering the lower bound of the one-sigma credibility intervals provides that it is placed beyond $N=46$ with probability 84\%.
This very wide interval suggests that the posterior distribution of the separation energies is perhaps not the most appropriate quantity to consider.
To this end, for each model, we  consider the probability $p_{ex}(Z,N)$ of the predicted separation energy $S_{1n/2n}^*(Z,N)$ to be positive under the posterior probability distribution conditioned on the experimental masses available. In the Bayesian paradigm, this probability is
$p_{ex}(Z,N):=p(S_{1n/2n}^*(Z,N) > 0 | S_{1n/2n})$.
The insert in Fig.~\ref{fig-Cachain} shows $p_{ex}$ for the Ca chain.
The model-averaged existence probabilities for the Ca region are shown in Fig.~\ref{fig-dripline}(a) assuming  uniform  prior weights. 
(For the values of $p_{ex}$ for individual models, see Sec.~III.C of SM.)
As noticed in Ref.\cite{Tarasov2018}, the  $N=35$ isotones $^{52}$Cl and $^{53}$Ar, as well as $^{49}$S represent a challenge for nuclear mass models. Our results in Fig.~\ref{fig-dripline}(a) confirm this finding through the low calculated prior{-average} $p_{ex}$ values for these nuclei.  Indeed, with the exception of SV-min, UNEDF0, and FRDM-2012, other models  calculate them to be either marginally bound or to lie outside the one-neutron drip line.  Since  
$^{49}$S, $^{52}$Cl and $^{53}$Ar do exist \cite{Tarasov2009,Tarasov2018}, this prior knowledge can inform the model averaging process \cite{Hoeting1999,Was00,Bernardo1994} through posterior weights: 
\begin{equation}\label{eq-post-weights}
w_k := p\left(\mathcal{M}_k | {}^{52}\text{Cl}, {}^{53}\text{Ar}, {}^{49}\text{S} \ \text{exist}\right)
\end{equation}
(see additional discussion in  SM). The weight $w_k$ reflects the ability of the model $\mathcal{M}_k$ to predict the existence of nuclei in the Ca region. In this respect UNEDF0+GP is superior, see Table S1 in SM.
We emphasize that conditioning with respect to these three nuclei corresponds actually to conditioning over the observed nuclei in the whole Ca  region, since  other experimentally-observed isotopes are predicted to be bound by the global  models considered.
The values of $p_{ex}$ obtained in this way are shown in Fig.~\ref{fig-dripline}(b).

As shown in Figs.~\ref{fig-Cachain} and \ref{fig-dripline}, the nucleus $^{68}$Ca is expected to be bound. However, as seen in Fig.~\ref{fig-Cachain}, $S_{2n}$ approaches zero very gradually; this results in a  spread of predictions of individual models. 
According to the average $p_{ex}$,
$^{61}$Ca and $^{71}$Ti are expected to be  $1n$-unstable while  the $2n$ drip line extends all the way to $^{72}$Ca and $^{78}$Ti. 
The nucleus $^{59}$K -- for which one event  was registered in Ref.~\cite{Tarasov2018} -- is expected to be firmly neutron-bound.
By comparing Figs.~\ref{fig-dripline}(a) and (b) one can immediately assess the impact of
the discovery of $^{52}$Cl,  $^{53}$Ar, and $^{49}$S  on drip-line predictions: the $2n$ drip line obtained with posterior weights generally extends by two neutron numbers for  odd-$Z$ chains.  

\textit{Conclusions --} 
In summary, in this Letter 
we quantified the neutron-stability of the nucleus in terms of its existence probability $p_{ex}$, i.e., the Bayesian posterior probability that the neutron separation energy is positive.
Our results are fairly consistent with recent experimental findings \cite{Tarasov2018}: $^{60}$Ca is expected to be well bound with $S_{2n}\approx 5$\,MeV while   $^{49}$S, 
$^{52}$Cl, and $^{53}$Ar are marginally-bound threshold systems. 

We emphasize that the nuclear model itself is not capable of gauging the likelihood of existence. To overcome this problem, we introduce a machine learning algorithm, with a stochastic exploration part and a deterministic modeling part, which, when combined, result in Bayesian statistical machine learning. One could say this is supervised learning, with the nuclear  modeling and the choice of priors representing two aspects of the supervision.
 
The Bayesian model averaging employed in this Letter is based on global DFT/mean-field models.  Therefore the  computed probabilities of existence
are conditional on the correctness of the DFT framework.
Currently, many $A$-body methods based on realistic inter-nucleon interactions calculate the two-neutron drip line at $^{60}$Ca. Since Bayesian machine learning requires
a sufficient number of data points to extrapolate with reasonable certainty,
$A$-body  models  are  not yet  amenable to statistical analysis   as the corresponding global mass tables are difficult to compute.  It will  be 
extremely valuable to apply a Bayesian uncertainty quantification analysis to $A$-body mass tables when those become available.

The extrapolation outcomes discussed in this Letter will be tested by  experimental data from rare-isotope facilities.  New mass measurements on neutron-rich nuclei will help to develop increasingly more quantitative models of the atomic nucleus and also allow for a higher-fidelity statistical analysis.  As illuminated by our Bayesian  analysis 
of  $^{49}$S, 
$^{52}$Cl, and $^{53}$Ar,  experimental discoveries of new nuclides will also be crucial for delineating the detailed behavior  of the nuclear mass surface, including the placement of particle drip lines.

\begin{acknowledgements}
Useful comments from Alexandra Gade and Heiko Hergert are gratefully appreciated. This work was supported by the U.S. Department of Energy under Award Numbers DE-SC0013365 (Office of Science), DE-SC0018083 (Office of Science, NUCLEI SciDAC-4 collaboration), and  DOE-DE-NA0002847 (NNSA, the Stewardship Science Academic Alliances program).
\end{acknowledgements}


\begin{thebibliography}{58}%
\makeatletter
\providecommand \@ifxundefined [1]{%
 \@ifx{#1\undefined}
}%
\providecommand \@ifnum [1]{%
 \ifnum #1\expandafter \@firstoftwo
 \else \expandafter \@secondoftwo
 \fi
}%
\providecommand \@ifx [1]{%
 \ifx #1\expandafter \@firstoftwo
 \else \expandafter \@secondoftwo
 \fi
}%
\providecommand \natexlab [1]{#1}%
\providecommand \enquote  [1]{``#1''}%
\providecommand \bibnamefont  [1]{#1}%
\providecommand \bibfnamefont [1]{#1}%
\providecommand \citenamefont [1]{#1}%
\providecommand \href@noop [0]{\@secondoftwo}%
\providecommand \href [0]{\begingroup \@sanitize@url \@href}%
\providecommand \@href[1]{\@@startlink{#1}\@@href}%
\providecommand \@@href[1]{\endgroup#1\@@endlink}%
\providecommand \@sanitize@url [0]{\catcode `\\12\catcode `\$12\catcode
  `\&12\catcode `\#12\catcode `\^12\catcode `\_12\catcode `\%12\relax}%
\providecommand \@@startlink[1]{}%
\providecommand \@@endlink[0]{}%
\providecommand \url  [0]{\begingroup\@sanitize@url \@url }%
\providecommand \@url [1]{\endgroup\@href {#1}{\urlprefix }}%
\providecommand \urlprefix  [0]{URL }%
\providecommand \Eprint [0]{\href }%
\providecommand \doibase [0]{http://dx.doi.org/}%
\providecommand \selectlanguage [0]{\@gobble}%
\providecommand \bibinfo  [0]{\@secondoftwo}%
\providecommand \bibfield  [0]{\@secondoftwo}%
\providecommand \translation [1]{[#1]}%
\providecommand \BibitemOpen [0]{}%
\providecommand \bibitemStop [0]{}%
\providecommand \bibitemNoStop [0]{.\EOS\space}%
\providecommand \EOS [0]{\spacefactor3000\relax}%
\providecommand \BibitemShut  [1]{\csname bibitem#1\endcsname}%
\let\auto@bib@innerbib\@empty
\bibitem [{\citenamefont {Tarasov}\ \emph {et~al.}(2018)\citenamefont {Tarasov}
  \emph {et~al.}}]{Tarasov2018}%
  \BibitemOpen
  \bibfield  {author} {\bibinfo {author} {\bibfnamefont {O.~B.}\ \bibnamefont
  {Tarasov}} \emph {et~al.},\ }\href {\doibase 10.1103/PhysRevLett.121.022501}
  {\bibfield  {journal} {\bibinfo  {journal} {Phys. Rev. Lett}\ }\textbf
  {\bibinfo {volume} {121}},\ \bibinfo {pages} {022501} (\bibinfo {year}
  {2018})}\BibitemShut {NoStop}%
\bibitem [{\citenamefont {Michimasa}\ \emph {et~al.}(2018)\citenamefont
  {Michimasa} \emph {et~al.}}]{Michimasa2018}%
  \BibitemOpen
  \bibfield  {author} {\bibinfo {author} {\bibfnamefont {S.}~\bibnamefont
  {Michimasa}} \emph {et~al.},\ }\href {\doibase
  10.1103/PhysRevLett.121.022506} {\bibfield  {journal} {\bibinfo  {journal}
  {Phys. Rev. Lett.}\ }\textbf {\bibinfo {volume} {121}},\ \bibinfo {pages}
  {022506} (\bibinfo {year} {2018})}\BibitemShut {NoStop}%
\bibitem [{\citenamefont {Thoennessen}(2016)}]{Thoennessen2016}%
  \BibitemOpen
  \bibfield  {author} {\bibinfo {author} {\bibfnamefont {M.}~\bibnamefont
  {Thoennessen}},\ }\href {https://www.springer.com/us/book/9783319317618}
  {\emph {\bibinfo {title} {The Discovery of Isotopes}}}\ (\bibinfo
  {publisher} {Springer},\ \bibinfo {year} {2016})\BibitemShut {NoStop}%
\bibitem [{\citenamefont {Erler}\ \emph {et~al.}(2012)\citenamefont {Erler},
  \citenamefont {Birge}, \citenamefont {Kortelainen}, \citenamefont
  {Nazarewicz}, \citenamefont {Olsen}, \citenamefont {Perhac},\ and\
  \citenamefont {Stoitsov}}]{Erl12a}%
  \BibitemOpen
  \bibfield  {author} {\bibinfo {author} {\bibfnamefont {J.}~\bibnamefont
  {Erler}}, \bibinfo {author} {\bibfnamefont {N.}~\bibnamefont {Birge}},
  \bibinfo {author} {\bibfnamefont {M.}~\bibnamefont {Kortelainen}}, \bibinfo
  {author} {\bibfnamefont {W.}~\bibnamefont {Nazarewicz}}, \bibinfo {author}
  {\bibfnamefont {E.}~\bibnamefont {Olsen}}, \bibinfo {author} {\bibfnamefont
  {A.}~\bibnamefont {Perhac}}, \ and\ \bibinfo {author} {\bibfnamefont
  {M.}~\bibnamefont {Stoitsov}},\ }\href {\doibase 10.1038/nature11188}
  {\bibfield  {journal} {\bibinfo  {journal} {Nature}\ }\textbf {\bibinfo
  {volume} {486}},\ \bibinfo {pages} {509} (\bibinfo {year}
  {2012})}\BibitemShut {NoStop}%
\bibitem [{\citenamefont {Agbemava}\ \emph {et~al.}(2014)\citenamefont
  {Agbemava}, \citenamefont {Afanasjev}, \citenamefont {Ray},\ and\
  \citenamefont {Ring}}]{Agbemava2014}%
  \BibitemOpen
  \bibfield  {author} {\bibinfo {author} {\bibfnamefont {S.~E.}\ \bibnamefont
  {Agbemava}}, \bibinfo {author} {\bibfnamefont {A.~V.}\ \bibnamefont
  {Afanasjev}}, \bibinfo {author} {\bibfnamefont {D.}~\bibnamefont {Ray}}, \
  and\ \bibinfo {author} {\bibfnamefont {P.}~\bibnamefont {Ring}},\ }\href
  {\doibase 10.1103/PhysRevC.89.054320} {\bibfield  {journal} {\bibinfo
  {journal} {Phys. Rev. C}\ }\textbf {\bibinfo {volume} {89}},\ \bibinfo
  {pages} {054320} (\bibinfo {year} {2014})}\BibitemShut {NoStop}%
\bibitem [{\citenamefont {Brink}\ and\ \citenamefont
  {Broglia}(2005)}]{Brink2005}%
  \BibitemOpen
  \bibfield  {author} {\bibinfo {author} {\bibfnamefont {D.~M.}\ \bibnamefont
  {Brink}}\ and\ \bibinfo {author} {\bibfnamefont {R.~A.}\ \bibnamefont
  {Broglia}},\ }\href
  {http://admin.cambridge.org/academic/subjects/physics/particle-physics-and-nuclear-physics/nuclear-superfluidity-pairing-finite-systems}
  {\emph {\bibinfo {title} {Nuclear Superfluidity, Pairing in Finite
  Systems}}}\ (\bibinfo  {publisher} {Cambridge University Press, Cambridge,
  UK},\ \bibinfo {year} {2005})\BibitemShut {NoStop}%
\bibitem [{\citenamefont {Tarasov}\ \emph {et~al.}(2009)\citenamefont {Tarasov}
  \emph {et~al.}}]{Tarasov2009}%
  \BibitemOpen
  \bibfield  {author} {\bibinfo {author} {\bibfnamefont {O.~B.}\ \bibnamefont
  {Tarasov}} \emph {et~al.},\ }\href {\doibase 10.1103/PhysRevLett.102.142501}
  {\bibfield  {journal} {\bibinfo  {journal} {Phys. Rev. Lett.}\ }\textbf
  {\bibinfo {volume} {102}},\ \bibinfo {pages} {142501} (\bibinfo {year}
  {2009})}\BibitemShut {NoStop}%
\bibitem [{\citenamefont {Tarasov}\ \emph {et~al.}(2013)\citenamefont {Tarasov}
  \emph {et~al.}}]{Tarasov2013}%
  \BibitemOpen
  \bibfield  {author} {\bibinfo {author} {\bibfnamefont {O.~B.}\ \bibnamefont
  {Tarasov}} \emph {et~al.},\ }\href {\doibase 10.1103/PhysRevC.87.054612}
  {\bibfield  {journal} {\bibinfo  {journal} {Phys. Rev. C}\ }\textbf {\bibinfo
  {volume} {87}},\ \bibinfo {pages} {054612} (\bibinfo {year}
  {2013})}\BibitemShut {NoStop}%
\bibitem [{\citenamefont {Leistenschneider}\ \emph {et~al.}(2018)\citenamefont
  {Leistenschneider} \emph {et~al.}}]{Leistenschneider2018}%
  \BibitemOpen
  \bibfield  {author} {\bibinfo {author} {\bibfnamefont {E.}~\bibnamefont
  {Leistenschneider}} \emph {et~al.},\ }\href {\doibase
  10.1103/PhysRevLett.120.062503} {\bibfield  {journal} {\bibinfo  {journal}
  {Phys. Rev. Lett.}\ }\textbf {\bibinfo {volume} {120}},\ \bibinfo {pages}
  {062503} (\bibinfo {year} {2018})}\BibitemShut {NoStop}%
\bibitem [{\citenamefont {Hagen}\ \emph {et~al.}(2014)\citenamefont {Hagen},
  \citenamefont {Papenbrock}, \citenamefont {Hjorth-Jensen},\ and\
  \citenamefont {Dean}}]{Hagen2014}%
  \BibitemOpen
  \bibfield  {author} {\bibinfo {author} {\bibfnamefont {G.}~\bibnamefont
  {Hagen}}, \bibinfo {author} {\bibfnamefont {T.}~\bibnamefont {Papenbrock}},
  \bibinfo {author} {\bibfnamefont {M.}~\bibnamefont {Hjorth-Jensen}}, \ and\
  \bibinfo {author} {\bibfnamefont {D.~J.}\ \bibnamefont {Dean}},\ }\href
  {\doibase 10.1088/0034-4885/77/9/096302} {\bibfield  {journal} {\bibinfo
  {journal} {Rep. Prog. Phys.}\ }\textbf {\bibinfo {volume} {77}},\ \bibinfo
  {pages} {096302} (\bibinfo {year} {2014})}\BibitemShut {NoStop}%
\bibitem [{\citenamefont {Hergert}\ \emph {et~al.}(2016)\citenamefont
  {Hergert}, \citenamefont {Bogner}, \citenamefont {Morris}, \citenamefont
  {Schwenk},\ and\ \citenamefont {Tsukiyama}}]{Hergert2016}%
  \BibitemOpen
  \bibfield  {author} {\bibinfo {author} {\bibfnamefont {H.}~\bibnamefont
  {Hergert}}, \bibinfo {author} {\bibfnamefont {S.}~\bibnamefont {Bogner}},
  \bibinfo {author} {\bibfnamefont {T.}~\bibnamefont {Morris}}, \bibinfo
  {author} {\bibfnamefont {A.}~\bibnamefont {Schwenk}}, \ and\ \bibinfo
  {author} {\bibfnamefont {K.}~\bibnamefont {Tsukiyama}},\ }\href {\doibase
  https://doi.org/10.1016/j.physrep.2015.12.007} {\bibfield  {journal}
  {\bibinfo  {journal} {Phys. Rep.}\ }\textbf {\bibinfo {volume} {621}},\
  \bibinfo {pages} {165 } (\bibinfo {year} {2016})}\BibitemShut {NoStop}%
\bibitem [{\citenamefont {Bender}\ \emph {et~al.}(2003)\citenamefont {Bender},
  \citenamefont {Heenen},\ and\ \citenamefont {Reinhard}}]{Ben03}%
  \BibitemOpen
  \bibfield  {author} {\bibinfo {author} {\bibfnamefont {M.}~\bibnamefont
  {Bender}}, \bibinfo {author} {\bibfnamefont {P.-H.}\ \bibnamefont {Heenen}},
  \ and\ \bibinfo {author} {\bibfnamefont {P.-G.}\ \bibnamefont {Reinhard}},\
  }\href {\doibase 10.1103/RevModPhys.75.121} {\bibfield  {journal} {\bibinfo
  {journal} {Rev. Mod. Phys.}\ }\textbf {\bibinfo {volume} {75}},\ \bibinfo
  {pages} {121} (\bibinfo {year} {2003})}\BibitemShut {NoStop}%
\bibitem [{\citenamefont {Forss{\'e}n}\ \emph {et~al.}(2013)\citenamefont
  {Forss{\'e}n}, \citenamefont {Hagen}, \citenamefont {Hjorth-Jensen},
  \citenamefont {Nazarewicz},\ and\ \citenamefont {Rotureau}}]{Forssen2013}%
  \BibitemOpen
  \bibfield  {author} {\bibinfo {author} {\bibfnamefont {C.}~\bibnamefont
  {Forss{\'e}n}}, \bibinfo {author} {\bibfnamefont {G.}~\bibnamefont {Hagen}},
  \bibinfo {author} {\bibfnamefont {M.}~\bibnamefont {Hjorth-Jensen}}, \bibinfo
  {author} {\bibfnamefont {W.}~\bibnamefont {Nazarewicz}}, \ and\ \bibinfo
  {author} {\bibfnamefont {J.}~\bibnamefont {Rotureau}},\ }\href {\doibase
  10.1088/0031-8949/2013/T152/014022} {\bibfield  {journal} {\bibinfo
  {journal} {Phys. Scripta}\ }\textbf {\bibinfo {volume} {2013}},\ \bibinfo
  {pages} {014022} (\bibinfo {year} {2013})}\BibitemShut {NoStop}%
\bibitem [{\citenamefont {Nazarewicz}(2016)}]{Nazarewicz2016}%
  \BibitemOpen
  \bibfield  {author} {\bibinfo {author} {\bibfnamefont {W.}~\bibnamefont
  {Nazarewicz}},\ }\href {\doibase 10.1088/0954-3899/43/4/044002} {\bibfield
  {journal} {\bibinfo  {journal} {J. Phys. G}\ }\textbf {\bibinfo {volume}
  {43}},\ \bibinfo {pages} {044002} (\bibinfo {year} {2016})}\BibitemShut
  {NoStop}%
\bibitem [{\citenamefont {Hagen}\ \emph {et~al.}(2012)\citenamefont {Hagen},
  \citenamefont {Hjorth-Jensen}, \citenamefont {Jansen}, \citenamefont
  {Machleidt},\ and\ \citenamefont {Papenbrock}}]{Hagen2012}%
  \BibitemOpen
  \bibfield  {author} {\bibinfo {author} {\bibfnamefont {G.}~\bibnamefont
  {Hagen}}, \bibinfo {author} {\bibfnamefont {M.}~\bibnamefont
  {Hjorth-Jensen}}, \bibinfo {author} {\bibfnamefont {G.~R.}\ \bibnamefont
  {Jansen}}, \bibinfo {author} {\bibfnamefont {R.}~\bibnamefont {Machleidt}}, \
  and\ \bibinfo {author} {\bibfnamefont {T.}~\bibnamefont {Papenbrock}},\
  }\href {\doibase 10.1103/PhysRevLett.109.032502} {\bibfield  {journal}
  {\bibinfo  {journal} {Phys. Rev. Lett.}\ }\textbf {\bibinfo {volume} {109}},\
  \bibinfo {pages} {032502} (\bibinfo {year} {2012})}\BibitemShut {NoStop}%
\bibitem [{\citenamefont {Holt}\ \emph {et~al.}(2014)\citenamefont {Holt},
  \citenamefont {Men\'endez}, \citenamefont {Simonis},\ and\ \citenamefont
  {Schwenk}}]{Holt2014}%
  \BibitemOpen
  \bibfield  {author} {\bibinfo {author} {\bibfnamefont {J.~D.}\ \bibnamefont
  {Holt}}, \bibinfo {author} {\bibfnamefont {J.}~\bibnamefont {Men\'endez}},
  \bibinfo {author} {\bibfnamefont {J.}~\bibnamefont {Simonis}}, \ and\
  \bibinfo {author} {\bibfnamefont {A.}~\bibnamefont {Schwenk}},\ }\href
  {\doibase 10.1103/PhysRevC.90.024312} {\bibfield  {journal} {\bibinfo
  {journal} {Phys. Rev. C}\ }\textbf {\bibinfo {volume} {90}},\ \bibinfo
  {pages} {024312} (\bibinfo {year} {2014})}\BibitemShut {NoStop}%
\bibitem [{\citenamefont {Som\`a}\ \emph {et~al.}(2014)\citenamefont {Som\`a},
  \citenamefont {Cipollone}, \citenamefont {Barbieri}, \citenamefont
  {Navr\'atil},\ and\ \citenamefont {Duguet}}]{Soma2014}%
  \BibitemOpen
  \bibfield  {author} {\bibinfo {author} {\bibfnamefont {V.}~\bibnamefont
  {Som\`a}}, \bibinfo {author} {\bibfnamefont {A.}~\bibnamefont {Cipollone}},
  \bibinfo {author} {\bibfnamefont {C.}~\bibnamefont {Barbieri}}, \bibinfo
  {author} {\bibfnamefont {P.}~\bibnamefont {Navr\'atil}}, \ and\ \bibinfo
  {author} {\bibfnamefont {T.}~\bibnamefont {Duguet}},\ }\href {\doibase
  10.1103/PhysRevC.89.061301} {\bibfield  {journal} {\bibinfo  {journal} {Phys.
  Rev. C}\ }\textbf {\bibinfo {volume} {89}},\ \bibinfo {pages} {061301}
  (\bibinfo {year} {2014})}\BibitemShut {NoStop}%
\bibitem [{\citenamefont {Hergert}\ \emph {et~al.}(2014)\citenamefont
  {Hergert}, \citenamefont {Bogner}, \citenamefont {Morris}, \citenamefont
  {Binder}, \citenamefont {Calci}, \citenamefont {Langhammer},\ and\
  \citenamefont {Roth}}]{Hergert2014}%
  \BibitemOpen
  \bibfield  {author} {\bibinfo {author} {\bibfnamefont {H.}~\bibnamefont
  {Hergert}}, \bibinfo {author} {\bibfnamefont {S.~K.}\ \bibnamefont {Bogner}},
  \bibinfo {author} {\bibfnamefont {T.~D.}\ \bibnamefont {Morris}}, \bibinfo
  {author} {\bibfnamefont {S.}~\bibnamefont {Binder}}, \bibinfo {author}
  {\bibfnamefont {A.}~\bibnamefont {Calci}}, \bibinfo {author} {\bibfnamefont
  {J.}~\bibnamefont {Langhammer}}, \ and\ \bibinfo {author} {\bibfnamefont
  {R.}~\bibnamefont {Roth}},\ }\href {\doibase 10.1103/PhysRevC.90.041302}
  {\bibfield  {journal} {\bibinfo  {journal} {Phys. Rev. C}\ }\textbf {\bibinfo
  {volume} {90}},\ \bibinfo {pages} {041302} (\bibinfo {year}
  {2014})}\BibitemShut {NoStop}%
\bibitem [{\citenamefont {Hagen}\ \emph {et~al.}(2015)\citenamefont {Hagen},
  \citenamefont {Ekstr\"{o}m}, \citenamefont {Forss\'{e}n}, \citenamefont
  {Jansen}, \citenamefont {Nazarewicz}, \citenamefont {Papenbrock},
  \citenamefont {Wendt}, \citenamefont {Bacca}, \citenamefont {Barnea},
  \citenamefont {Carlsson}, \citenamefont {Drischler}, \citenamefont {Hebeler},
  \citenamefont {Hjorth-Jensen}, \citenamefont {Miorelli}, \citenamefont
  {Orlandini}, \citenamefont {Schwenk},\ and\ \citenamefont
  {Simonis}}]{Hagen2015}%
  \BibitemOpen
  \bibfield  {author} {\bibinfo {author} {\bibfnamefont {G.}~\bibnamefont
  {Hagen}}, \bibinfo {author} {\bibfnamefont {A.}~\bibnamefont {Ekstr\"{o}m}},
  \bibinfo {author} {\bibfnamefont {C.}~\bibnamefont {Forss\'{e}n}}, \bibinfo
  {author} {\bibfnamefont {G.~R.}\ \bibnamefont {Jansen}}, \bibinfo {author}
  {\bibfnamefont {W.}~\bibnamefont {Nazarewicz}}, \bibinfo {author}
  {\bibfnamefont {T.}~\bibnamefont {Papenbrock}}, \bibinfo {author}
  {\bibfnamefont {K.~A.}\ \bibnamefont {Wendt}}, \bibinfo {author}
  {\bibfnamefont {S.}~\bibnamefont {Bacca}}, \bibinfo {author} {\bibfnamefont
  {N.}~\bibnamefont {Barnea}}, \bibinfo {author} {\bibfnamefont
  {B.}~\bibnamefont {Carlsson}}, \bibinfo {author} {\bibfnamefont
  {C.}~\bibnamefont {Drischler}}, \bibinfo {author} {\bibfnamefont
  {K.}~\bibnamefont {Hebeler}}, \bibinfo {author} {\bibfnamefont
  {M.}~\bibnamefont {Hjorth-Jensen}}, \bibinfo {author} {\bibfnamefont
  {M.}~\bibnamefont {Miorelli}}, \bibinfo {author} {\bibfnamefont
  {G.}~\bibnamefont {Orlandini}}, \bibinfo {author} {\bibfnamefont
  {A.}~\bibnamefont {Schwenk}}, \ and\ \bibinfo {author} {\bibfnamefont
  {J.}~\bibnamefont {Simonis}},\ }\href {\doibase 10.1038/nphys3529} {\bibfield
   {journal} {\bibinfo  {journal} {Nat. Phys.}\ }\textbf {\bibinfo {volume}
  {12}},\ \bibinfo {pages} {186} (\bibinfo {year} {2015})}\BibitemShut
  {NoStop}%
\bibitem [{\citenamefont {{Garcia Ruiz}}\ \emph {et~al.}(2016)\citenamefont
  {{Garcia Ruiz}} \emph {et~al.}}]{GarciaRuiz16}%
  \BibitemOpen
  \bibfield  {author} {\bibinfo {author} {\bibfnamefont {R.~F.}\ \bibnamefont
  {{Garcia Ruiz}}} \emph {et~al.},\ }\href {\doibase 10.1038/NPHYS3645}
  {\bibfield  {journal} {\bibinfo  {journal} {Nat. Phys.}\ }\textbf {\bibinfo
  {volume} {12}},\ \bibinfo {pages} {594} (\bibinfo {year} {2016})}\BibitemShut
  {NoStop}%
\bibitem [{\citenamefont {Simonis}\ \emph {et~al.}(2017)\citenamefont
  {Simonis}, \citenamefont {Stroberg}, \citenamefont {Hebeler}, \citenamefont
  {Holt},\ and\ \citenamefont {Schwenk}}]{Simonis2017}%
  \BibitemOpen
  \bibfield  {author} {\bibinfo {author} {\bibfnamefont {J.}~\bibnamefont
  {Simonis}}, \bibinfo {author} {\bibfnamefont {S.~R.}\ \bibnamefont
  {Stroberg}}, \bibinfo {author} {\bibfnamefont {K.}~\bibnamefont {Hebeler}},
  \bibinfo {author} {\bibfnamefont {J.~D.}\ \bibnamefont {Holt}}, \ and\
  \bibinfo {author} {\bibfnamefont {A.}~\bibnamefont {Schwenk}},\ }\href
  {\doibase 10.1103/PhysRevC.96.014303} {\bibfield  {journal} {\bibinfo
  {journal} {Phys. Rev. C}\ }\textbf {\bibinfo {volume} {96}},\ \bibinfo
  {pages} {014303} (\bibinfo {year} {2017})}\BibitemShut {NoStop}%
\bibitem [{\citenamefont {Stroberg}\ \emph {et~al.}(2017)\citenamefont
  {Stroberg}, \citenamefont {Calci}, \citenamefont {Hergert}, \citenamefont
  {Holt}, \citenamefont {Bogner}, \citenamefont {Roth},\ and\ \citenamefont
  {Schwenk}}]{Stroberg2017}%
  \BibitemOpen
  \bibfield  {author} {\bibinfo {author} {\bibfnamefont {S.~R.}\ \bibnamefont
  {Stroberg}}, \bibinfo {author} {\bibfnamefont {A.}~\bibnamefont {Calci}},
  \bibinfo {author} {\bibfnamefont {H.}~\bibnamefont {Hergert}}, \bibinfo
  {author} {\bibfnamefont {J.~D.}\ \bibnamefont {Holt}}, \bibinfo {author}
  {\bibfnamefont {S.~K.}\ \bibnamefont {Bogner}}, \bibinfo {author}
  {\bibfnamefont {R.}~\bibnamefont {Roth}}, \ and\ \bibinfo {author}
  {\bibfnamefont {A.}~\bibnamefont {Schwenk}},\ }\href {\doibase
  10.1103/PhysRevLett.118.032502} {\bibfield  {journal} {\bibinfo  {journal}
  {Phys. Rev. Lett.}\ }\textbf {\bibinfo {volume} {118}},\ \bibinfo {pages}
  {032502} (\bibinfo {year} {2017})}\BibitemShut {NoStop}%
\bibitem [{\citenamefont {Hagen}\ \emph {et~al.}(2013)\citenamefont {Hagen},
  \citenamefont {Hagen}, \citenamefont {Hammer},\ and\ \citenamefont
  {Platter}}]{Hagen2013}%
  \BibitemOpen
  \bibfield  {author} {\bibinfo {author} {\bibfnamefont {G.}~\bibnamefont
  {Hagen}}, \bibinfo {author} {\bibfnamefont {P.}~\bibnamefont {Hagen}},
  \bibinfo {author} {\bibfnamefont {H.-W.}\ \bibnamefont {Hammer}}, \ and\
  \bibinfo {author} {\bibfnamefont {L.}~\bibnamefont {Platter}},\ }\href
  {\doibase 10.1103/PhysRevLett.111.132501} {\bibfield  {journal} {\bibinfo
  {journal} {Phys. Rev. Lett.}\ }\textbf {\bibinfo {volume} {111}},\ \bibinfo
  {pages} {132501} (\bibinfo {year} {2013})}\BibitemShut {NoStop}%
\bibitem [{\citenamefont {Neufcourt}\ \emph {et~al.}(2018)\citenamefont
  {Neufcourt}, \citenamefont {Cao}, \citenamefont {Nazarewicz},\ and\
  \citenamefont {Viens}}]{Neufcourt2018}%
  \BibitemOpen
  \bibfield  {author} {\bibinfo {author} {\bibfnamefont {L.}~\bibnamefont
  {Neufcourt}}, \bibinfo {author} {\bibfnamefont {Y.}~\bibnamefont {Cao}},
  \bibinfo {author} {\bibfnamefont {W.}~\bibnamefont {Nazarewicz}}, \ and\
  \bibinfo {author} {\bibfnamefont {F.}~\bibnamefont {Viens}},\ }\href
  {\doibase 10.1103/PhysRevC.98.034318} {\bibfield  {journal} {\bibinfo
  {journal} {Phys. Rev. C}\ }\textbf {\bibinfo {volume} {98}},\ \bibinfo
  {pages} {034318} (\bibinfo {year} {2018})}\BibitemShut {NoStop}%
\bibitem [{\citenamefont {Bartel}\ \emph {et~al.}(1982)\citenamefont {Bartel},
  \citenamefont {Quentin}, \citenamefont {Brack}, \citenamefont {Guet},\ and\
  \citenamefont {H{\aa}kansson}}]{Bartel1982}%
  \BibitemOpen
  \bibfield  {author} {\bibinfo {author} {\bibfnamefont {J.}~\bibnamefont
  {Bartel}}, \bibinfo {author} {\bibfnamefont {P.}~\bibnamefont {Quentin}},
  \bibinfo {author} {\bibfnamefont {M.}~\bibnamefont {Brack}}, \bibinfo
  {author} {\bibfnamefont {C.}~\bibnamefont {Guet}}, \ and\ \bibinfo {author}
  {\bibfnamefont {H.-B.}\ \bibnamefont {H{\aa}kansson}},\ }\href {\doibase
  10.1016/0375-9474(82)90403-1} {\bibfield  {journal} {\bibinfo  {journal}
  {Nucl. Phys. A}\ }\textbf {\bibinfo {volume} {386}},\ \bibinfo {pages} {79 }
  (\bibinfo {year} {1982})}\BibitemShut {NoStop}%
\bibitem [{\citenamefont {Dobaczewski}\ \emph {et~al.}(1984)\citenamefont
  {Dobaczewski}, \citenamefont {Flocard},\ and\ \citenamefont
  {Treiner}}]{Dob84}%
  \BibitemOpen
  \bibfield  {author} {\bibinfo {author} {\bibfnamefont {J.}~\bibnamefont
  {Dobaczewski}}, \bibinfo {author} {\bibfnamefont {H.}~\bibnamefont
  {Flocard}}, \ and\ \bibinfo {author} {\bibfnamefont {J.}~\bibnamefont
  {Treiner}},\ }\href {\doibase 10.1016/0375-9474(84)90433-0} {\bibfield
  {journal} {\bibinfo  {journal} {Nucl. Phys. A}\ }\textbf {\bibinfo {volume}
  {422}},\ \bibinfo {pages} {103 } (\bibinfo {year} {1984})}\BibitemShut
  {NoStop}%
\bibitem [{\citenamefont {Chabanat}\ \emph {et~al.}(1995)\citenamefont
  {Chabanat}, \citenamefont {Bonche}, \citenamefont {Haensel}, \citenamefont
  {Meyer},\ and\ \citenamefont {Schaeffer}}]{Chabanat1995}%
  \BibitemOpen
  \bibfield  {author} {\bibinfo {author} {\bibfnamefont {E.}~\bibnamefont
  {Chabanat}}, \bibinfo {author} {\bibfnamefont {P.}~\bibnamefont {Bonche}},
  \bibinfo {author} {\bibfnamefont {P.}~\bibnamefont {Haensel}}, \bibinfo
  {author} {\bibfnamefont {J.}~\bibnamefont {Meyer}}, \ and\ \bibinfo {author}
  {\bibfnamefont {R.}~\bibnamefont {Schaeffer}},\ }\href
  {http://stacks.iop.org/1402-4896/1995/i=T56/a=034} {\bibfield  {journal}
  {\bibinfo  {journal} {Physica Scr.}\ }\textbf {\bibinfo {volume} {1995}},\
  \bibinfo {pages} {231} (\bibinfo {year} {1995})}\BibitemShut {NoStop}%
\bibitem [{\citenamefont {Kl{\"{u}}pfel}\ \emph {et~al.}(2009)\citenamefont
  {Kl{\"{u}}pfel}, \citenamefont {Reinhard}, \citenamefont {B{\"{u}}rvenich},\
  and\ \citenamefont {Maruhn}}]{Kluepfel2009}%
  \BibitemOpen
  \bibfield  {author} {\bibinfo {author} {\bibfnamefont {P.}~\bibnamefont
  {Kl{\"{u}}pfel}}, \bibinfo {author} {\bibfnamefont {P.-G.}\ \bibnamefont
  {Reinhard}}, \bibinfo {author} {\bibfnamefont {T.~J.}\ \bibnamefont
  {B{\"{u}}rvenich}}, \ and\ \bibinfo {author} {\bibfnamefont {J.~A.}\
  \bibnamefont {Maruhn}},\ }\href {\doibase 10.1103/PhysRevC.79.034310}
  {\bibfield  {journal} {\bibinfo  {journal} {Phys. Rev. C}\ }\textbf {\bibinfo
  {volume} {79}},\ \bibinfo {pages} {034310} (\bibinfo {year}
  {2009})}\BibitemShut {NoStop}%
\bibitem [{\citenamefont {Kortelainen}\ \emph {et~al.}(2010)\citenamefont
  {Kortelainen}, \citenamefont {Lesinski}, \citenamefont {Mor\'e},
  \citenamefont {Nazarewicz}, \citenamefont {Sarich}, \citenamefont {Schunck},
  \citenamefont {Stoitsov},\ and\ \citenamefont {Wild}}]{UNEDF0}%
  \BibitemOpen
  \bibfield  {author} {\bibinfo {author} {\bibfnamefont {M.}~\bibnamefont
  {Kortelainen}}, \bibinfo {author} {\bibfnamefont {T.}~\bibnamefont
  {Lesinski}}, \bibinfo {author} {\bibfnamefont {J.}~\bibnamefont {Mor\'e}},
  \bibinfo {author} {\bibfnamefont {W.}~\bibnamefont {Nazarewicz}}, \bibinfo
  {author} {\bibfnamefont {J.}~\bibnamefont {Sarich}}, \bibinfo {author}
  {\bibfnamefont {N.}~\bibnamefont {Schunck}}, \bibinfo {author} {\bibfnamefont
  {M.~V.}\ \bibnamefont {Stoitsov}}, \ and\ \bibinfo {author} {\bibfnamefont
  {S.}~\bibnamefont {Wild}},\ }\href {\doibase 10.1103/PhysRevC.82.024313}
  {\bibfield  {journal} {\bibinfo  {journal} {Phys. Rev. C}\ }\textbf {\bibinfo
  {volume} {82}},\ \bibinfo {pages} {024313} (\bibinfo {year}
  {2010})}\BibitemShut {NoStop}%
\bibitem [{\citenamefont {Kortelainen}\ \emph {et~al.}(2012)\citenamefont
  {Kortelainen}, \citenamefont {McDonnell}, \citenamefont {Nazarewicz},
  \citenamefont {Reinhard}, \citenamefont {Sarich}, \citenamefont {Schunck},
  \citenamefont {Stoitsov},\ and\ \citenamefont {Wild}}]{UNEDF1}%
  \BibitemOpen
  \bibfield  {author} {\bibinfo {author} {\bibfnamefont {M.}~\bibnamefont
  {Kortelainen}}, \bibinfo {author} {\bibfnamefont {J.}~\bibnamefont
  {McDonnell}}, \bibinfo {author} {\bibfnamefont {W.}~\bibnamefont
  {Nazarewicz}}, \bibinfo {author} {\bibfnamefont {P.-G.}\ \bibnamefont
  {Reinhard}}, \bibinfo {author} {\bibfnamefont {J.}~\bibnamefont {Sarich}},
  \bibinfo {author} {\bibfnamefont {N.}~\bibnamefont {Schunck}}, \bibinfo
  {author} {\bibfnamefont {M.~V.}\ \bibnamefont {Stoitsov}}, \ and\ \bibinfo
  {author} {\bibfnamefont {S.~M.}\ \bibnamefont {Wild}},\ }\href {\doibase
  10.1103/PhysRevC.85.024304} {\bibfield  {journal} {\bibinfo  {journal} {Phys.
  Rev. C}\ }\textbf {\bibinfo {volume} {85}},\ \bibinfo {pages} {024304}
  (\bibinfo {year} {2012})}\BibitemShut {NoStop}%
\bibitem [{mas()}]{massexplorer}%
  \BibitemOpen
  \href@noop {} {}\bibinfo {note} {Mass Explorer,
  \url{http://massexplorer.frib.msu.edu/}}\BibitemShut {NoStop}%
\bibitem [{\citenamefont {Kortelainen}\ \emph {et~al.}(2014)\citenamefont
  {Kortelainen}, \citenamefont {McDonnell}, \citenamefont {Nazarewicz},
  \citenamefont {Olsen}, \citenamefont {Reinhard}, \citenamefont {Sarich},
  \citenamefont {Schunck}, \citenamefont {Wild}, \citenamefont {Davesne},
  \citenamefont {Erler},\ and\ \citenamefont {Pastore}}]{UNEDF2}%
  \BibitemOpen
  \bibfield  {author} {\bibinfo {author} {\bibfnamefont {M.}~\bibnamefont
  {Kortelainen}}, \bibinfo {author} {\bibfnamefont {J.}~\bibnamefont
  {McDonnell}}, \bibinfo {author} {\bibfnamefont {W.}~\bibnamefont
  {Nazarewicz}}, \bibinfo {author} {\bibfnamefont {E.}~\bibnamefont {Olsen}},
  \bibinfo {author} {\bibfnamefont {P.-G.}\ \bibnamefont {Reinhard}}, \bibinfo
  {author} {\bibfnamefont {J.}~\bibnamefont {Sarich}}, \bibinfo {author}
  {\bibfnamefont {N.}~\bibnamefont {Schunck}}, \bibinfo {author} {\bibfnamefont
  {S.~M.}\ \bibnamefont {Wild}}, \bibinfo {author} {\bibfnamefont
  {D.}~\bibnamefont {Davesne}}, \bibinfo {author} {\bibfnamefont
  {J.}~\bibnamefont {Erler}}, \ and\ \bibinfo {author} {\bibfnamefont
  {A.}~\bibnamefont {Pastore}},\ }\href {\doibase 10.1103/PhysRevC.89.054314}
  {\bibfield  {journal} {\bibinfo  {journal} {Phys. Rev. C}\ }\textbf {\bibinfo
  {volume} {89}},\ \bibinfo {pages} {054314} (\bibinfo {year}
  {2014})}\BibitemShut {NoStop}%
\bibitem [{\citenamefont {M{\"o}ller}\ \emph {et~al.}(2016)\citenamefont
  {M{\"o}ller}, \citenamefont {Sierk}, \citenamefont {Ichikawa},\ and\
  \citenamefont {Sagawa}}]{Moller2012}%
  \BibitemOpen
  \bibfield  {author} {\bibinfo {author} {\bibfnamefont {P.}~\bibnamefont
  {M{\"o}ller}}, \bibinfo {author} {\bibfnamefont {A.}~\bibnamefont {Sierk}},
  \bibinfo {author} {\bibfnamefont {T.}~\bibnamefont {Ichikawa}}, \ and\
  \bibinfo {author} {\bibfnamefont {H.}~\bibnamefont {Sagawa}},\ }\href
  {\doibase 10.1016/j.adt.2015.10.002} {\bibfield  {journal} {\bibinfo
  {journal} {At. Data Nucl. Data Tables}\ }\textbf {\bibinfo {volume}
  {109-110}},\ \bibinfo {pages} {1 } (\bibinfo {year} {2016})}\BibitemShut
  {NoStop}%
\bibitem [{\citenamefont {Goriely}\ \emph {et~al.}(2013)\citenamefont
  {Goriely}, \citenamefont {Chamel},\ and\ \citenamefont
  {Pearson}}]{Goriely2013}%
  \BibitemOpen
  \bibfield  {author} {\bibinfo {author} {\bibfnamefont {S.}~\bibnamefont
  {Goriely}}, \bibinfo {author} {\bibfnamefont {N.}~\bibnamefont {Chamel}}, \
  and\ \bibinfo {author} {\bibfnamefont {J.~M.}\ \bibnamefont {Pearson}},\
  }\href {\doibase 10.1103/PhysRevC.88.024308} {\bibfield  {journal} {\bibinfo
  {journal} {Phys. Rev. C}\ }\textbf {\bibinfo {volume} {88}},\ \bibinfo
  {pages} {024308} (\bibinfo {year} {2013})}\BibitemShut {NoStop}%
\bibitem [{\citenamefont {Bonneau}\ \emph {et~al.}(2007)\citenamefont
  {Bonneau}, \citenamefont {Quentin},\ and\ \citenamefont
  {M\"oller}}]{Bonneau2007}%
  \BibitemOpen
  \bibfield  {author} {\bibinfo {author} {\bibfnamefont {L.}~\bibnamefont
  {Bonneau}}, \bibinfo {author} {\bibfnamefont {P.}~\bibnamefont {Quentin}}, \
  and\ \bibinfo {author} {\bibfnamefont {P.}~\bibnamefont {M\"oller}},\ }\href
  {\doibase 10.1103/PhysRevC.76.024320} {\bibfield  {journal} {\bibinfo
  {journal} {Phys. Rev. C}\ }\textbf {\bibinfo {volume} {76}},\ \bibinfo
  {pages} {024320} (\bibinfo {year} {2007})}\BibitemShut {NoStop}%
\bibitem [{\citenamefont {Schunck}\ \emph {et~al.}(2010)\citenamefont
  {Schunck}, \citenamefont {Dobaczewski}, \citenamefont {McDonnell},
  \citenamefont {Mor\'e}, \citenamefont {Nazarewicz}, \citenamefont {Sarich},\
  and\ \citenamefont {Stoitsov}}]{Schunck2010}%
  \BibitemOpen
  \bibfield  {author} {\bibinfo {author} {\bibfnamefont {N.}~\bibnamefont
  {Schunck}}, \bibinfo {author} {\bibfnamefont {J.}~\bibnamefont
  {Dobaczewski}}, \bibinfo {author} {\bibfnamefont {J.}~\bibnamefont
  {McDonnell}}, \bibinfo {author} {\bibfnamefont {J.}~\bibnamefont {Mor\'e}},
  \bibinfo {author} {\bibfnamefont {W.}~\bibnamefont {Nazarewicz}}, \bibinfo
  {author} {\bibfnamefont {J.}~\bibnamefont {Sarich}}, \ and\ \bibinfo {author}
  {\bibfnamefont {M.~V.}\ \bibnamefont {Stoitsov}},\ }\href {\doibase
  10.1103/PhysRevC.81.024316} {\bibfield  {journal} {\bibinfo  {journal} {Phys.
  Rev. C}\ }\textbf {\bibinfo {volume} {81}},\ \bibinfo {pages} {024316}
  (\bibinfo {year} {2010})}\BibitemShut {NoStop}%
\bibitem [{\citenamefont {Afanasjev}(2015)}]{Afanasjev2015a}%
  \BibitemOpen
  \bibfield  {author} {\bibinfo {author} {\bibfnamefont {A.~V.}\ \bibnamefont
  {Afanasjev}},\ }\href {http://stacks.iop.org/0954-3899/42/i=3/a=034002}
  {\bibfield  {journal} {\bibinfo  {journal} {J. Phys. G}\ }\textbf {\bibinfo
  {volume} {42}},\ \bibinfo {pages} {034002} (\bibinfo {year}
  {2015})}\BibitemShut {NoStop}%
\bibitem [{\citenamefont {Dobaczewski}\ and\ \citenamefont
  {Nazarewicz}(2013)}]{Dob2013}%
  \BibitemOpen
  \bibfield  {author} {\bibinfo {author} {\bibfnamefont {J.}~\bibnamefont
  {Dobaczewski}}\ and\ \bibinfo {author} {\bibfnamefont {W.}~\bibnamefont
  {Nazarewicz}},\ }\enquote {\bibinfo {title} {{Hartree-Fock-Bogoliubov}
  solution of the pairing hamiltonian in finite nuclei},}\ in\ \href {\doibase
  10.1142/9789814412490_0004} {\emph {\bibinfo {booktitle} {Fifty Years of
  Nuclear BCS}}}\ (\bibinfo  {publisher} {World Scientific},\ \bibinfo {year}
  {2013})\ pp.\ \bibinfo {pages} {40--60}\BibitemShut {NoStop}%
\bibitem [{\citenamefont {Audi}\ \emph {et~al.}(2003)\citenamefont {Audi},
  \citenamefont {Wapstra},\ and\ \citenamefont {Thibault}}]{AME03b}%
  \BibitemOpen
  \bibfield  {author} {\bibinfo {author} {\bibfnamefont {G.}~\bibnamefont
  {Audi}}, \bibinfo {author} {\bibfnamefont {A.}~\bibnamefont {Wapstra}}, \
  and\ \bibinfo {author} {\bibfnamefont {C.}~\bibnamefont {Thibault}},\ }\href
  {\doibase 10.1016/j.nuclphysa.2003.11.003} {\bibfield  {journal} {\bibinfo
  {journal} {Nucl. Phys. A}\ }\textbf {\bibinfo {volume} {729}},\ \bibinfo
  {pages} {337 } (\bibinfo {year} {2003})}\BibitemShut {NoStop}%
\bibitem [{\citenamefont {{Wang}}\ \emph {et~al.}(2017)\citenamefont {{Wang}},
  \citenamefont {{Audi}}, \citenamefont {{Kondev}}, \citenamefont {{Huang}},
  \citenamefont {{Naimi}},\ and\ \citenamefont {{Xu}}}]{AME16b}%
  \BibitemOpen
  \bibfield  {author} {\bibinfo {author} {\bibfnamefont {M.}~\bibnamefont
  {{Wang}}}, \bibinfo {author} {\bibfnamefont {G.}~\bibnamefont {{Audi}}},
  \bibinfo {author} {\bibfnamefont {F.~G.}\ \bibnamefont {{Kondev}}}, \bibinfo
  {author} {\bibfnamefont {W.~J.}\ \bibnamefont {{Huang}}}, \bibinfo {author}
  {\bibfnamefont {S.}~\bibnamefont {{Naimi}}}, \ and\ \bibinfo {author}
  {\bibfnamefont {X.}~\bibnamefont {{Xu}}},\ }\href {\doibase
  10.1088/1674-1137/41/3/030003} {\bibfield  {journal} {\bibinfo  {journal}
  {Chin. Phys. C}\ }\textbf {\bibinfo {volume} {41}},\ \bibinfo {eid} {030003}
  (\bibinfo {year} {2017})}\BibitemShut {NoStop}%
\bibitem [{\citenamefont {Rasmussen}\ and\ \citenamefont
  {Williams}(2006)}]{RasmussenWilliams}%
  \BibitemOpen
  \bibfield  {author} {\bibinfo {author} {\bibfnamefont {C.~E.}\ \bibnamefont
  {Rasmussen}}\ and\ \bibinfo {author} {\bibfnamefont {C.~K.~I.}\ \bibnamefont
  {Williams}},\ }\href {www.GaussianProcess.org/gpml} {\emph {\bibinfo {title}
  {Gaussian Processes for Machine Learning}}}\ (\bibinfo  {publisher} {MIT
  Press},\ \bibinfo {year} {2006})\BibitemShut {NoStop}%
\bibitem [{\citenamefont {MacKay}(2005)}]{MacKay}%
  \BibitemOpen
  \bibfield  {author} {\bibinfo {author} {\bibfnamefont {D.}~\bibnamefont
  {MacKay}},\ }\href {http://www.inference.org.uk/itprnn/book.pdf} {\emph
  {\bibinfo {title} {Information Theory, Inference, and Learning Algorithms}}}\
  (\bibinfo  {publisher} {Cambridge University Press},\ \bibinfo {year}
  {2005})\BibitemShut {NoStop}%
\bibitem [{SM()}]{SM}%
  \BibitemOpen
  \href@noop {} {}\bibinfo {note} {See Supplemental Material at
  \url{http://link.aps.org/supplemental/XXX}}\BibitemShut {NoStop}%
\bibitem [{\citenamefont {Hastings}(1970)}]{Hastings1970}%
  \BibitemOpen
  \bibfield  {author} {\bibinfo {author} {\bibfnamefont {W.~K.}\ \bibnamefont
  {Hastings}},\ }\href {\doibase 10.1093/biomet/57.1.97} {\bibfield  {journal}
  {\bibinfo  {journal} {Biometrika}\ }\textbf {\bibinfo {volume} {57}},\
  \bibinfo {pages} {97} (\bibinfo {year} {1970})}\BibitemShut {NoStop}%
\bibitem [{\citenamefont {Utama}\ and\ \citenamefont
  {Piekarewicz}(2018)}]{Utama18}%
  \BibitemOpen
  \bibfield  {author} {\bibinfo {author} {\bibfnamefont {R.}~\bibnamefont
  {Utama}}\ and\ \bibinfo {author} {\bibfnamefont {J.}~\bibnamefont
  {Piekarewicz}},\ }\href {\doibase 10.1103/PhysRevC.97.014306} {\bibfield
  {journal} {\bibinfo  {journal} {Phys. Rev. C}\ }\textbf {\bibinfo {volume}
  {97}},\ \bibinfo {pages} {014306} (\bibinfo {year} {2018})}\BibitemShut
  {NoStop}%
\bibitem [{\citenamefont {Athanassopoulos}\ \emph {et~al.}(2004)\citenamefont
  {Athanassopoulos}, \citenamefont {Mavrommatis}, \citenamefont {Gernoth},\
  and\ \citenamefont {Clark}}]{Athanassopoulos2004}%
  \BibitemOpen
  \bibfield  {author} {\bibinfo {author} {\bibfnamefont {S.}~\bibnamefont
  {Athanassopoulos}}, \bibinfo {author} {\bibfnamefont {E.}~\bibnamefont
  {Mavrommatis}}, \bibinfo {author} {\bibfnamefont {K.}~\bibnamefont
  {Gernoth}}, \ and\ \bibinfo {author} {\bibfnamefont {J.}~\bibnamefont
  {Clark}},\ }\href {\doibase 10.1016/j.nuclphysa.2004.08.006} {\bibfield
  {journal} {\bibinfo  {journal} {Nucl. Phys. A}\ }\textbf {\bibinfo {volume}
  {743}},\ \bibinfo {pages} {222 } (\bibinfo {year} {2004})}\BibitemShut
  {NoStop}%
\bibitem [{\citenamefont {Bayram}\ and\ \citenamefont
  {Akkoyun}(2017)}]{Bayram2017}%
  \BibitemOpen
  \bibfield  {author} {\bibinfo {author} {\bibfnamefont {T.}~\bibnamefont
  {Bayram}}\ and\ \bibinfo {author} {\bibfnamefont {S.}~\bibnamefont
  {Akkoyun}},\ }\href {\doibase 10.1051/epjconf/201714612033} {\bibfield
  {journal} {\bibinfo  {journal} {EPJ Web Conf.}\ }\textbf {\bibinfo {volume}
  {146}},\ \bibinfo {pages} {12033} (\bibinfo {year} {2017})}\BibitemShut
  {NoStop}%
\bibitem [{\citenamefont {Yuan}(2016)}]{Yuan2016}%
  \BibitemOpen
  \bibfield  {author} {\bibinfo {author} {\bibfnamefont {C.}~\bibnamefont
  {Yuan}},\ }\href {\doibase 10.1103/PhysRevC.93.034310} {\bibfield  {journal}
  {\bibinfo  {journal} {Phys. Rev. C}\ }\textbf {\bibinfo {volume} {93}},\
  \bibinfo {pages} {034310} (\bibinfo {year} {2016})}\BibitemShut {NoStop}%
\bibitem [{\citenamefont {Utama}\ \emph {et~al.}(2016)\citenamefont {Utama},
  \citenamefont {Piekarewicz},\ and\ \citenamefont {Prosper}}]{Utama16}%
  \BibitemOpen
  \bibfield  {author} {\bibinfo {author} {\bibfnamefont {R.}~\bibnamefont
  {Utama}}, \bibinfo {author} {\bibfnamefont {J.}~\bibnamefont {Piekarewicz}},
  \ and\ \bibinfo {author} {\bibfnamefont {H.~B.}\ \bibnamefont {Prosper}},\
  }\href {\doibase 10.1103/PhysRevC.93.014311} {\bibfield  {journal} {\bibinfo
  {journal} {Phys. Rev. C}\ }\textbf {\bibinfo {volume} {93}},\ \bibinfo
  {pages} {014311} (\bibinfo {year} {2016})}\BibitemShut {NoStop}%
\bibitem [{\citenamefont {Utama}\ and\ \citenamefont
  {Piekarewicz}(2017)}]{Utama17}%
  \BibitemOpen
  \bibfield  {author} {\bibinfo {author} {\bibfnamefont {R.}~\bibnamefont
  {Utama}}\ and\ \bibinfo {author} {\bibfnamefont {J.}~\bibnamefont
  {Piekarewicz}},\ }\href {\doibase 10.1103/PhysRevC.96.044308} {\bibfield
  {journal} {\bibinfo  {journal} {Phys. Rev. C}\ }\textbf {\bibinfo {volume}
  {96}},\ \bibinfo {pages} {044308} (\bibinfo {year} {2017})}\BibitemShut
  {NoStop}%
\bibitem [{\citenamefont {Bertsch}\ and\ \citenamefont
  {Bingham}(2017)}]{Bertsch2017}%
  \BibitemOpen
  \bibfield  {author} {\bibinfo {author} {\bibfnamefont {G.~F.}\ \bibnamefont
  {Bertsch}}\ and\ \bibinfo {author} {\bibfnamefont {D.}~\bibnamefont
  {Bingham}},\ }\href {\doibase 10.1103/PhysRevLett.119.252501} {\bibfield
  {journal} {\bibinfo  {journal} {Phys. Rev. Lett.}\ }\textbf {\bibinfo
  {volume} {119}},\ \bibinfo {pages} {252501} (\bibinfo {year}
  {2017})}\BibitemShut {NoStop}%
\bibitem [{\citenamefont {Zhang}\ \emph {et~al.}(2017)\citenamefont {Zhang},
  \citenamefont {Wang}, \citenamefont {Yin}, \citenamefont {Chen},\ and\
  \citenamefont {Zhang}}]{Zhang2017}%
  \BibitemOpen
  \bibfield  {author} {\bibinfo {author} {\bibfnamefont {H.~F.}\ \bibnamefont
  {Zhang}}, \bibinfo {author} {\bibfnamefont {L.~H.}\ \bibnamefont {Wang}},
  \bibinfo {author} {\bibfnamefont {J.~P.}\ \bibnamefont {Yin}}, \bibinfo
  {author} {\bibfnamefont {P.~H.}\ \bibnamefont {Chen}}, \ and\ \bibinfo
  {author} {\bibfnamefont {H.~F.}\ \bibnamefont {Zhang}},\ }\href
  {http://stacks.iop.org/0954-3899/44/i=4/a=045110} {\bibfield  {journal}
  {\bibinfo  {journal} {J. Phys. G}\ }\textbf {\bibinfo {volume} {44}},\
  \bibinfo {pages} {045110} (\bibinfo {year} {2017})}\BibitemShut {NoStop}%
\bibitem [{\citenamefont {Niu}\ and\ \citenamefont {Liang}(2018)}]{Niu2018}%
  \BibitemOpen
  \bibfield  {author} {\bibinfo {author} {\bibfnamefont {Z.}~\bibnamefont
  {Niu}}\ and\ \bibinfo {author} {\bibfnamefont {H.}~\bibnamefont {Liang}},\
  }\href {\doibase 10.1016/j.physletb.2018.01.002} {\bibfield  {journal}
  {\bibinfo  {journal} {Phys. Lett. B}\ }\textbf {\bibinfo {volume} {778}},\
  \bibinfo {pages} {48 } (\bibinfo {year} {2018})}\BibitemShut {NoStop}%
\bibitem [{\citenamefont {Gneiting}\ and\ \citenamefont
  {Raftery}(2007)}]{Gneiting2007}%
  \BibitemOpen
  \bibfield  {author} {\bibinfo {author} {\bibfnamefont {T.}~\bibnamefont
  {Gneiting}}\ and\ \bibinfo {author} {\bibfnamefont {A.~E.}\ \bibnamefont
  {Raftery}},\ }\href {\doibase 10.1198/016214506000001437} {\bibfield
  {journal} {\bibinfo  {journal} {J. Amer. Statist. Assoc.}\ }\textbf {\bibinfo
  {volume} {102}},\ \bibinfo {pages} {359} (\bibinfo {year}
  {2007})}\BibitemShut {NoStop}%
\bibitem [{\citenamefont {Gneiting}\ \emph {et~al.}(2007)\citenamefont
  {Gneiting}, \citenamefont {Balabdaoui},\ and\ \citenamefont
  {Raftery}}]{Gneiting2007a}%
  \BibitemOpen
  \bibfield  {author} {\bibinfo {author} {\bibfnamefont {T.}~\bibnamefont
  {Gneiting}}, \bibinfo {author} {\bibfnamefont {F.}~\bibnamefont
  {Balabdaoui}}, \ and\ \bibinfo {author} {\bibfnamefont {A.~E.}\ \bibnamefont
  {Raftery}},\ }\href {\doibase 10.1111/j.1467-9868.2007.00587.x} {\bibfield
  {journal} {\bibinfo  {journal} {J. Roy. Stat. Soc. Ser. B Stat. Methodol.}\
  }\textbf {\bibinfo {volume} {69}},\ \bibinfo {pages} {243} (\bibinfo {year}
  {2007})}\BibitemShut {NoStop}%
\bibitem [{\citenamefont {Hoeting}\ \emph {et~al.}(1999)\citenamefont
  {Hoeting}, \citenamefont {Madigan}, \citenamefont {Raftery},\ and\
  \citenamefont {Volinsky}}]{Hoeting1999}%
  \BibitemOpen
  \bibfield  {author} {\bibinfo {author} {\bibfnamefont {J.~A.}\ \bibnamefont
  {Hoeting}}, \bibinfo {author} {\bibfnamefont {D.}~\bibnamefont {Madigan}},
  \bibinfo {author} {\bibfnamefont {A.~E.}\ \bibnamefont {Raftery}}, \ and\
  \bibinfo {author} {\bibfnamefont {C.~T.}\ \bibnamefont {Volinsky}},\ }\href
  {\doibase 10.1214/ss/1009212519} {\bibfield  {journal} {\bibinfo  {journal}
  {Statist. Sci.}\ }\textbf {\bibinfo {volume} {14}},\ \bibinfo {pages} {382}
  (\bibinfo {year} {1999})}\BibitemShut {NoStop}%
\bibitem [{\citenamefont {Wasserman}(2000)}]{Was00}%
  \BibitemOpen
  \bibfield  {author} {\bibinfo {author} {\bibfnamefont {L.}~\bibnamefont
  {Wasserman}},\ }\href {\doibase 10.1006/jmps.1999.1278} {\bibfield  {journal}
  {\bibinfo  {journal} {J. Math. Psych.}\ }\textbf {\bibinfo {volume} {44}},\
  \bibinfo {pages} {92 } (\bibinfo {year} {2000})}\BibitemShut {NoStop}%
\bibitem [{\citenamefont {Bernardo}\ and\ \citenamefont
  {Smith}(1994)}]{Bernardo1994}%
  \BibitemOpen
  \bibfield  {author} {\bibinfo {author} {\bibfnamefont {J.~M.}\ \bibnamefont
  {Bernardo}}\ and\ \bibinfo {author} {\bibfnamefont {A.~F.~M.}\ \bibnamefont
  {Smith}},\ }\enquote {\bibinfo {title} {Reference analysis},}\ in\ \href
  {\doibase 10.1002/9780470316870} {\emph {\bibinfo {booktitle} {Bayesian
  Theory}}}\ (\bibinfo  {publisher} {Wiley},\ \bibinfo {year} {1994})\ Chap.\
  \bibinfo {chapter} {Inference}\BibitemShut {NoStop}%
\end{thebibliography}
%

\end{document}